\documentclass[letterpaper, 10 pt, conference]{ieeeconf}  

\usepackage{amsmath,amssymb,amsfonts}
\usepackage{graphicx}
\usepackage{textcomp}
\usepackage{xcolor}
\def\BibTeX{{\rm B\kern-.05em{\sc i\kern-.025em b}\kern-.08em    T\kern-.1667em\lower.7ex\hbox{E}\kern-.125emX}}
\usepackage{verbatim}
\usepackage{paralist}
\usepackage{graphics} 
\usepackage{epsfig} 
\usepackage{mathptmx} 
\usepackage{times} 
\usepackage{physics}
\usepackage{mdwmath}
\usepackage{mdwtab}
\usepackage[hidelinks]{hyperref}
\usepackage{algpseudocode}
\usepackage{algorithm}
\usepackage{caption}
\usepackage{subcaption}
\usepackage{cite}

\newtheorem{remark}{Remark}
\usepackage{tikz}
\usetikzlibrary{shapes, arrows.meta, positioning}

\IEEEoverridecommandlockouts                              
\title{\LARGE \bf
Robust Entanglement Generation in Bipartite Quantum Systems Using Optimal Control
}

\author{Nahid Binandeh Dehaghani, A. Pedro Aguiar, Rafal Wisniewski
\thanks{N. Dehaghani and R. Wisniewski are with Department of Electronic Systems, Aalborg University, Fredrik Bajers vej 7c, DK-9220 Aalborg, Denmark
 {\tt\small nahidbd@es.aau.dk, raf@es.aau.dk}}
\thanks{A. Aguiar is with the Research Center for Systems and Technologies (SYSTEC-ARISE), Faculty of Engineering, University of Porto, Rua Dr. Roberto Frias sn, i219, 4200-465 Porto, Portugal
        {\tt\small pedro.aguiar@fe.up.pt}}%
 \thanks{The authors acknowledge the support of the Danish e-Infrastructure Consortium (DeiC) and the National Quantum Algorithm Academy (NQAA) through the Postdoctoral Scholarship 
 under the project "Quantum-Driven Solutions for Multi-Agent Systems and Advanced Computation". This work was also partially supported by UID/00147- Research Center for Systems and Technologies (SYSTEC) - and the Associate Laboratory Advanced Production and Intelligent Systems (ARISE, 10.54499/LA/P/0112/2020) funded by Fundação para a Ciência e a Tecnologia, I.P./ MCTES through the national funds.}
 }

\begin{document}

\maketitle
\thispagestyle{empty}
\pagestyle{empty}

\begin{abstract}
Quantum entanglement is a key resource for quantum technologies, yet its efficient and high-fidelity generation remains a challenge due to the complexity of quantum dynamics. This paper presents a quantum optimal control framework to maximize bipartite entanglement within a fixed time horizon, under bounded control inputs. By leveraging Pontryagin’s Minimum Principle, we derive a set of necessary conditions that guide the design of time-dependent control fields to steer a two-qubit system toward maximally entangled Bell states. The entanglement is quantified using concurrence, and the control objective is formulated as maximizing this measure at the terminal time. Our approach is validated through numerical simulations of Liouville–von Neumann dynamics. The results demonstrate the effectiveness of switching-based control strategies in achieving robust entanglement, offering insights into practical implementations of quantum control for entanglement generation in quantum networks.
\end{abstract}

\section{Introduction}
\begin{figure}[h!]
    \centering
    \includegraphics[width=0.5\textwidth]{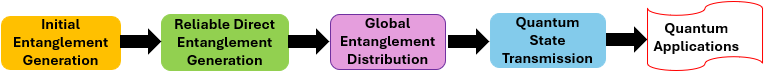} 
    \caption{A quantum network workflow illustrating the progression from initial entanglement generation (physical layer) to quantum applications. The workflow includes reliable direct entanglement generation, global entanglement distribution, and quantum state transmission, enabling practical quantum-enhanced applications.}
    \label{fig:quantum_network_workflow}
\end{figure}

Quantum entanglement \cite{horodecki2009quantum} is a physical phenomenon in which the quantum states of particles within a system are intrinsically linked, such that the state of each particle cannot be characterized independently of the others, regardless of the spatial separation between them. This concept lies at the core of the difference between classical and quantum physics, as entanglement represents a fundamental aspect of quantum mechanics absent in classical frameworks. The unique properties of entanglement play a central role in quantum computing, communication, and sensing, where it functions as a fundamental resource for processing and transmitting information \cite{steane1998quantum}.
Notably, entanglement is a fundamental element in quantum networks, as experimentally demonstrated in~\cite{pompili2022experimental} and depicted in Fig.~\ref{fig:quantum_network_workflow}. It serves as the cornerstone for distributed quantum protocols, including teleportation, secure key distribution, and quantum-enhanced metrology. 

However, harnessing entanglement for practical applications requires the ability to reliably generate high-fidelity entangled states—a task complicated by the inherent complexity and fragility of quantum systems. Effective control techniques are therefore essential to manage these challenges and enable the robust implementation of entanglement-based technologies. 
Although various strategies for generating entanglement have been proposed and experimentally demonstrated \cite{Lee2022GenerationOM, kues2017chip, riera2023remotely, gonzalez2015generation, leifer2003optimal, zeng2023controlled, mabrok2014robust, platzer2010optimal}, the application of optimal control theory \cite{boscain2021introduction, koch2022quantum} to this problem has not been extensively studied. Quantum optimal control \cite{dehaghani2023quantum, dehaghani2022quantum, dehaghani2023quantumbalancing} provides a powerful framework for designing control strategies that guide quantum systems toward specific target states, such as maximally entangled configurations. 
By shaping appropriate time-dependent control fields, quantum optimal control enables both the enhancement of quantum system performance and the efficient generation of entanglement. Importantly, this framework allows for the formulation of optimization problems that balance competing objectives—such as maximizing entanglement fidelity while minimizing control resources.

In this work, we present a quantum optimal control framework for the efficient generation of entangled states. In this approach, we consider a closed quantum system and formulate an optimal control problem. The goal is to maximize entanglement within a 
fixed time interval, under bounded control inputs. In particular, we show that optimally designed control fields can effectively steer 
an initial perturbed separable state
toward maximally entangled Bell states. We start by reviewing some of the fundamental mathematical concepts of quantum entanglement, and then propose the quantum control problem considering a bipartite system. We numerically verify this approach for a two-qubit system evolving under Liouville–von Neumann dynamics, using a combination of drift and control Hamiltonians.
The results demonstrate the ability of the control strategy to robustly generate high levels of entanglement across different initial conditions, offering insights into practical control design for quantum technologies.


\noindent\textbf{Notation:} 
Throughout this work, we adopt Dirac notation to represent quantum states. A quantum state vector is denoted by \( |\psi\rangle \), and its corresponding dual vector, or bra, is written as \( \langle \psi| = (|\psi\rangle)^\dagger \). The tensor product between vectors or operators is indicated by the symbol \( \otimes \). 
Commutators between two operators \( A \) and \( B \) are denoted by \( [A, B] = AB - BA \). The trace of a square matrix \( A \) is represented by \( \operatorname{Tr}(A) \), and the partial trace over subsystem \( B \) is written as \( \operatorname{Tr}_B(\cdot) \). We use
$i$ to denote the imaginary unit, and $I$ to denote the identity matrix.
Finally, we denote the set of complex numbers by \( \mathbb{C} \), and the set of purely imaginary numbers by \( i\mathbb{R} \).

\section{Mathematical Foundations of Quantum Entanglement}
This section provides a structured overview of the mathematical foundations related to entanglement in bipartite quantum systems based on the concepts discussed in \cite{liintroduction, horodecki2001entanglement, eisert1999comparison, vidal2000entanglement}. We begin by introducing the formal representation of bipartite systems and the Schmidt decomposition. We then discuss the notions of separability and maximally entangled states, followed by a quantitative approach for entanglement, which captures the strength of quantum correlations.

\subsection{Bipartite System}
A bipartite system consists of two distinct subsystems, typically referred to as \(A\) and \(B\), which are often conceptualized as being spatially separated and associated with two different parties, commonly named Alice and Bob. The overall Hilbert space of the bipartite system is represented by the tensor product \( \mathbb{H}_A \otimes \mathbb{H}_B \), where \( \mathbb{H}_A \) and \( \mathbb{H}_B \) denote the Hilbert spaces corresponding to Alice's and Bob's subsystems, respectively.

An important structural feature of bipartite pure states is the \textit{Schmidt Decomposition}. This decomposition provides a simplified representation of any pure state in the form:
\begin{equation}\label{schmith}
|\psi\rangle_{AB} = \sum_j \lambda_j \, |e_j\rangle_A \otimes |e_j\rangle_B,
\end{equation}
where \( \{|e_j\rangle_A\} \) and \( \{|e_j\rangle_B\} \) are orthonormal basis for the subsystems \( \mathbb{H}_A \) and \( \mathbb{H}_B \), and \( \lambda_j \) are non-negative real coefficients known as Schmidt coefficients, satisfying the normalization condition \( \sum_j \lambda_j^2 = 1 \).
The number of non-zero Schmidt coefficients determines the \textit{Schmidt number} of the pure state.

\subsection{Separability and Maximally Entangled States}
The concept of entanglement is fundamentally linked to the notion of separability in quantum states. For a bipartite pure state \( |\psi\rangle_{AB} \), entanglement arises if and only if the state is non-separable, which is equivalent to having a Schmidt number greater than one. In the case of separable states, the pure state of the composite system in the Hilbert space \( \mathbb{H}_A \otimes \mathbb{H}_B \) can be expressed as a tensor product of individual pure states, where \eqref{schmith} reduces to
\begin{equation}\label{psisep}
 |\psi\rangle_{AB} = |\psi\rangle_A \otimes |\psi\rangle_B.   \end{equation}
In terms of maximally entangled states, a well-known example is the EPR pair (Einstein-Podolsky-Rosen pair), represented as:
\begin{equation}\label{bellphi+}
|\phi^+\rangle = \frac{1}{\sqrt{2}} (|00\rangle + |11\rangle).
 \end{equation}
This state has a Schmidt number of 2 and cannot be written as a simple product of two independent states as in \eqref{psisep}. It is considered maximally entangled because tracing out one subsystem (e.g., qubit \( B \)) leaves the other (qubit \( A \)) in a completely mixed state, which is not pure. The other three Bell states share this property of maximal entanglement and can be transformed into each other through local unitary operations, and are expressed as:
\begin{equation}\label{bells}
    \begin{aligned}
 |\phi^-\rangle = \frac{1}{\sqrt{2}} (|00\rangle - |11\rangle), \quad
|\psi^+\rangle &= \frac{1}{\sqrt{2}} (|01\rangle + |10\rangle), \\
|\psi^-\rangle = \frac{1}{\sqrt{2}} (|01\rangle - |10\rangle).
\end{aligned}
\end{equation}

The existence of maximally entangled states in bipartite systems is particularly advantageous, as they serve as benchmarks for comparing entanglement levels in other states. Extending the idea of maximally entangled states to higher-dimensional systems, the generalized form for two \( d \)-dimensional subsystems (often referred to as qudits) is expressed as:
\[
|\phi^+_d\rangle = \frac{1}{\sqrt{d}} (|00\rangle + |11\rangle + \dots + |d-1, d-1\rangle).
\]

\subsection{Entanglement Quantification Using Concurrence}
To quantify the entanglement of the pure state \( |\psi\rangle_{AB}\), we employ the \textit{concurrence entanglement measure} \( E_c(\rho) \), 
a standard and effective tool for quantifying quantum correlations.
The concurrence takes a value of zero for separable states and reaches its maximum value of one for maximally entangled states, such as Bell states  (e.g., \eqref{bellphi+}). For pure states in arbitrary dimensions, a generalized expression for concurrence entanglement measure is given by:
\begin{equation}\label{concurrence}
E_c(\rho) = \sqrt{2\left(1 - \operatorname{Tr}(\rho_M^2)\right)},
\end{equation}
where \( \rho_M \) is the reduced density matrix corresponding to a chosen subsystem \( M \), obtained by tracing out its complement. This formulation quantifies the entanglement between \( M \) and the rest of the system. 
In the case of a bipartite pure state \( |\psi\rangle_{AB} \), this reduces to the standard setting where the joint density operator is given by \( \rho_{AB} = |\psi\rangle_{AB} \langle \psi|_{AB}\). The reduced density operator for subsystem \( A \), denoted \( \rho_A \), is then obtained by tracing out subsystem \( B \):  \( \rho_A = \operatorname{Tr}_B \left( \rho_{AB} \right) \)
By using the Schmidt decomposition of \( \rho_{AB} \), the reduced state of subsystem \( A \) can be expressed as:
\begin{equation}\label{rhoA}
\rho_A = \sum_{j} \left( {I}_A \otimes \langle e_j |_B  \right) \rho_{AB} \left( {I}_A \otimes | e_j \rangle_B \right),
\end{equation}
where \({I}_A \) is the identity operator on \( \mathbb{H}_A \). This form is particularly useful when the global state \( \rho_{AB} \) is represented in matrix form, as it facilitates numerical computation of the reduced density matrix and its associated purity.

\section{Quantum Optimal Control for Entanglement Manipulation}\label{sec:optimal_control}
Having recalled the mathematical foundations of quantum entanglement, we now turn our attention to the use of optimal control theory for the purposeful generation and manipulation of entangled states. In this section, we formulate an optimal control problem aimed at enhancing entanglement, and we derive the necessary conditions for optimality using Pontryagin’s Minimum Principle (PMP). This framework enables systematic design of control inputs that drive the quantum system toward desired entangled configurations under physical and dynamical constraints.

We consider a bipartite quantum system composed of two interacting subsystems. The goal is to steer the system’s state toward a highly entangled final configuration in a fixed time interval. To achieve this, we employ the framework of quantum optimal control and apply PMP to determine the time-dependent external control fields that govern the system’s evolution. The objective is to generate entanglement efficiently. Specifically, we formulate an optimal control problem that seeks to maximize the concurrence of the reduced density matrix of subsystem \(A\)
within a fixed time horizon. This defines a fixed final-time optimal control problem, where the cost functional is:
\begin{equation}\label{cost}
\text{Minimize} \quad  J[\rho_A(T)] =-E_c(\rho_A(T)),
\end{equation}
where \( T \) is the fixed final time of the evolution.
The system evolves according to the Liouville–von Neumann equation:
\begin{equation}\label{liou}
\dot{\rho}(t) = -i[H(t), \rho(t)], \quad t \in [t_i, T]
\end{equation}
where \( H(t) = H_d + H_c(t) \) denotes the total Hamiltonian of the system, composed of a time-independent drift Hamiltonian \( H_d \) and a time-dependent control Hamiltonian \( H_c(t) \). The control term is given by
$H_c(t) = \sum_{k=1}^{m} u_k(t) H_k$,
where \( H_k \) are fixed control operators and \( u_k(t) \) are real-valued, time-dependent control inputs. The system evolves over a finite time horizon \( t \in [t_i, T] \), where \( t_i \) is the fixed initial time. 

Since we are interested in the entanglement between the two subsystems, control is applied indirectly through the evolution of the reduced density matrix \(\rho_A(t)\), obtained via the partial trace over subsystem \(B\):
\begin{equation}\label{liouA}
\dot \rho_A(t) = \operatorname{Tr}_B(\dot \rho(t)).
\end{equation}
This yields the effective dynamics for subsystem \(A\) that are targeted for optimization via the control field \(u(t)\), with the aim of maximizing concurrence at the final time. Equation \eqref{liouA} can further be analyzed using the Schmidt decomposition explained earlier in \eqref{rhoA}. 
Specifically, we write:
\begin{equation}\label{liouA2}
\dot{\rho}_A (t) = -i \sum_{j} \left( {I}_A \otimes \langle e_j|_B \right) [H(t), \rho(t)] \left( {I}_A \otimes |e_j \rangle_B \right),
\end{equation}
This decomposition allows the dynamics of \( \rho_A \) to be expressed in terms of the full system evolution and facilitates the computation of its functional derivative.

The initial condition of the system is specified by \( \rho(t_i) = \rho_i \), where \( \rho_i \in \mathbb{H}_\rho \) is the given initial density operator of the composite system at time \( t_i \). Here, \( \mathbb{H}_\rho \) denotes the set of admissible density operators defined on the Hilbert space \( \mathbb{H}_A \otimes \mathbb{H}_B \), that is,
\[
\mathbb{H}_\rho := \left\{ \rho \in \mathbb{C}^{d \times d} \,\middle|\, \rho = \rho^\dagger,\ \rho \geq 0,\ \operatorname{Tr}(\rho) = 1 \right\},
\]
where \( d = \dim(\mathbb{H}_A \otimes \mathbb{H}_B) \). 
From \( \rho_i \), the reduced density matrix at the initial time is obtained by tracing out subsystem \( B \), so \( \rho_A(t_i) = \operatorname{Tr}_B(\rho_i) \)

The control input \( u(t) \) belongs to the admissible set \( \mathcal{U} \):
\begin{equation}\label{u}
    u(t) \in \mathcal{U} := \left\{ u \in L^\infty([t_i, T]; \mathbb{R}^m) : \|u(t)\|_\infty \le u_{\max}\right\}
\end{equation} which consists of all essentially bounded, measurable functions \( u: [t_i, t_f] \rightarrow \mathbb{R}^m \) whose components satisfy the pointwise constraint \( \|u(t)\|_\infty \le u_{\max} \) for all \( t \in [t_i, T] \). This ensures that the control fields remain within physically realizable amplitude limits throughout the evolution.

In what follows, we derive the necessary conditions that characterize the optimal solution to the entanglement generation problem, based on equations \eqref{cost}, \eqref{liouA2}, and \eqref{u}, using PMP. 
The objective is to determine the optimal control trajectory \( u^*(t) \), the corresponding system state \( \rho^*(t) \), and the adjoint variable \( \lambda^*(t) \), such that the system evolves toward a highly entangled state in a fixed time interval, subject to the physical constraints of the quantum system.

To begin, we define the Pontryagin Hamiltonian as
\[
\mathcal{H}(\rho(t), u(t), \lambda(t), t) = \operatorname{Tr}\left(\lambda^\dagger(t)\, \dot{\rho}_A(t)\right), \quad t \in [t_i, T],
\]
where \( \lambda(t) \) is the adjoint variable associated with the reduced state \( \rho_A(t) \).
Substituting the Liouville–von Neumann equation into the Hamiltonian gives:
\begin{equation}\label{pontryagin}
\mathcal{H}(\rho(t), u(t), \lambda(t), t) = -i\, \operatorname{Tr} \left( \lambda^\dagger(t) \operatorname{Tr}_B \left( [H(t), \rho(t)] \right) \right).
\end{equation}
Since both \( H(t) \) and \( \rho(t) \) are Hermitian operators, the commutator \( [H, \rho] \) is skew-Hermitian. This follows from the general identity that for two generic Hermitian operators \( A \) and \( B \), the commutator \( [A, B] = A B-BA\) satisfies:
\[
[A, B]^\dagger = (AB - BA)^\dagger = B ^\dagger A^\dagger - A^\dagger B^\dagger = -[A, B].
\]
Hence, \( [H, \rho]^\dagger = -[H, \rho] \), confirming that \( [H, \rho] \) is skew-Hermitian.

As a direct consequence, all diagonal elements of a skew-Hermitian matrix are purely imaginary. Specifically, for any diagonal entry \( a_{ii} \), the skew-Hermitian condition implies \( a_{ii}^* = -a_{ii} \), which holds if and only if \( a_{ii} \in i\mathbb{R} \). 
Using the cyclic property of the trace, one concludes that
$$\operatorname{Tr}([H, \rho]) = \operatorname{Tr}(H \rho - \rho H) = \operatorname{Tr}(H \rho) - \operatorname{Tr}(\rho H) = 0.$$
Therefore, not only is the global commutator \( [H, \rho] \) traceless, but so is its partial trace over subsystem \( B \), 
$\operatorname{Tr}(\operatorname{Tr}_B([H, \rho])) = 0$.
This property is particularly useful when analyzing the reduced dynamics and constructing the Pontryagin Hamiltonian, since it constrains the form of the partial trace \( \operatorname{Tr}_B([H, \rho]) \) to a traceless skew-Hermitian matrix.

As established, the commutator \( [H(t), \rho(t)] \) is skew-Hermitian and traceless, and so is its partial trace over subsystem \( B \). For a bipartite system in which subsystem \( A \) is two-dimensional, this structure constrains the partial trace \( \operatorname{Tr}_B([H(t), \rho(t)]) \) to the space of \( 2 \times 2 \) traceless skew-Hermitian matrices. As such, the general form of this matrix can be expressed as:
    $\operatorname{Tr}_B([H(t), \rho(t)]) =  \left(
\begin{smallmatrix} \gamma_{11} & \gamma_{12} \\ -\gamma_{12}^* & -\gamma_{11} \end{smallmatrix}
\right)$,
where \( \gamma_{11} \in i\mathbb{R} \) is purely imaginary, and \( \gamma_{12} \in \mathbb{C} \) is a complex-valued off-diagonal entry. This canonical representation captures both the skew-Hermitian nature and the zero-trace condition of the partial commutator. It will be particularly useful when analyzing the structure of the Pontryagin Hamiltonian and deriving the associated optimal control laws. By defining the adjoint variable as
$\left(
\begin{smallmatrix} \lambda_{11} & \lambda_{12} \\ \lambda_{12}^* & \lambda_{22} \end{smallmatrix}
\right)$, one attains 
\begin{equation*}
\lambda^\dagger(t) \operatorname{Tr}_B \left( [H(t), \rho(t)]\right) =\left(
\begin{smallmatrix}
\lambda_{11} \gamma_{11} - \lambda_{12} \gamma_{12}^* & \lambda_{11} \gamma_{12} - \lambda_{12} \gamma_{11} \\
\lambda_{12}^* \gamma_{11} - \lambda_{22} \gamma_{12}^* & \lambda_{12}^* \gamma_{12} - \lambda_{22} \gamma_{11}
\end{smallmatrix}
\right) 
\end{equation*}
Given the formula in \eqref{pontryagin}, the Pontryagin Hamiltonian becomes:
\begin{align*}
\mathcal{H} 
&= -i \left( \lambda_{11} \gamma_{11} - \lambda_{12} \gamma_{12}^* + \lambda_{12}^* \gamma_{12} - \lambda_{22} \gamma_{11} \right) \\
&= -i \left( (\lambda_{11} - \lambda_{22}) \gamma_{11} + \lambda_{12}^* \gamma_{12} - (\lambda_{12}^* \gamma_{12})^* \right) \\
&= -i \left( (\lambda_{11} - \lambda_{22}) \gamma_{11} + 2i\, \Im(\lambda_{12}^* \gamma_{12}) \right).
\end{align*}
We note that \( \lambda_{11}, \lambda_{22} \in \mathbb{R} \) and \( \Im(\lambda_{12}^* \gamma_{12}) \in \mathbb{R} \), while \( \gamma_{11} \in i\mathbb{R} \). Therefore, the
term $\left( (\lambda_{11} - \lambda_{22}) \gamma_{11} + 2i\, \Im(\lambda_{12}^* \gamma_{12}) \right)$ is purely imaginary. As a result, multiplying this term by \( -i \) yields a real number. Hence, we conclude that the Pontryagin Hamiltonian \( \mathcal{H} \) is real-valued. This property is consistent with the physical interpretation of the Hamiltonian as a scalar performance measure in the optimal control framework.

Having established that the Pontryagin Hamiltonian \( \mathcal{H} \) is real-valued, we now proceed to derive the optimal control strategy. The control \( u^*(t) \) is determined by minimizing the Hamiltonian over the admissible control set \( \mathcal{U} \) at each instant of time \( t \). The optimality condition for minimization of Hamiltonian is expressed as follows:
\[
u^*(t) \in \arg\min_{u \in \mathcal{U}} \mathcal{H}(\rho^*(t), u, \lambda(t), t) \quad \text{for a.e.  } t \in [t_i, T].
\]
dependence of \( \mathcal{H} \) on the control inputs \( u_k(t) \), this minimization drives the optimal solution to lie on the boundary of the admissible set. Specifically, the control enters through terms of the form 
$-i\, u_k(t)\, \operatorname{Tr}\left( \lambda^\dagger(t) \operatorname{Tr}_B\left([H_k, \rho(t)]\right) \right),$
which results in a bang-bang type control law:
\[
u_k^*(t) = u_{k,\max} \cdot \text{sgn} \left( i\, \operatorname{Tr} \left( \lambda^\dagger(t)\, \operatorname{Tr}_B \left( [H_k, \rho(t)] \right) \right) \right),
\]
where 
$\text{sgn}(x) =
\begin{cases}
+1, & \text{if } x> 0 \\
-1, & \text{if } x \leq 0.
\end{cases}.$
This result implies that the optimal control switches between its maximum and minimum allowable values depending on the sign of the switching function, thereby exhibiting classical bang-bang behavior.
Next, we characterize the evolution of the adjoint variable \( \lambda(t) \), which plays a crucial role in the application of PMP. The adjoint evolves backward in time according to the following differential equation: 
\begin{align}
\dot{\lambda}(t) &= -\frac{\partial \mathcal{H}(\rho^*(t), u^*(t), \lambda(t), t) }{\partial \rho_A(t)} = -\lambda^\dagger(t)\, \dot{\rho}_A'(t) \label{adj}
\end{align}
Here, \( \dot{\rho}_A'(t) \) denotes the functional derivative of the reduced dynamics \( \dot{\rho}_A \) with respect to the reduced density matrix \( \rho_A \).
To compute \( \dot{\rho}_A'(t) \), we consider the decomposition of the full system’s density matrix \( \rho(t) \) into separable and correlated components. 
In the absence of quantum correlations, the system is assumed to be separable such that \( \rho(t) = \rho_A(t) \otimes \rho_B(t) \). In this case, the derivative of the total density matrix with respect to \( \rho_A \) becomes
$\frac{\partial \rho}{\partial \rho_A} = {I}_B \otimes \rho_B$,
which leads to the simplified expression for the functional derivative of the reduced dynamics:
\[
\dot{\rho}_A' = -i \sum_{j} \left( {I}_A \otimes \langle e_j |_B \right) \left[ H, {I}_B \otimes \rho_B \right] \left({I}_A \otimes |e_j\rangle _B \right).
\]
However, in the more general case where correlations exist between the subsystems, the total density matrix takes the form \( \rho = \rho_A\otimes \rho_B+ \rho_{\mathrm{c}} \). The derivative then becomes
$\frac{\partial \rho}{\partial \rho_A} = {I}_B \otimes \rho_B + \frac{\partial \rho_{\mathrm{c}}}{\partial \rho_A}$,
and the corresponding functional derivative of the reduced dynamics is given by:
\begin{equation*}
\begin{aligned}
\dot{\rho}_A'(t) &= -i \sum_{j}^r \left( {I}_A \otimes \langle e_j |_B \right) \left[ H, {I}_B \otimes \rho_B + \frac{\partial \rho_{\mathrm{c}}}{\partial \rho_A} \right] \left( {I}_A \otimes |e_j \rangle _B\right).
\end{aligned}
\end{equation*}
The term concerning $\rho_{\mathrm{c}}$ accounts for non-separable quantum correlations and may significantly influence the dynamics, especially in strongly entangled systems.

Since the final time $T$ is fixed and the terminal state is free, a transversality condition must be applied to our fixed-final-time optimal control problem. Specifically, at the final time $T$, the adjoint variable satisfies the boundary condition:
\begin{equation}\label{adjfinal}
\lambda(T) = \nabla_{\rho_A} J(\rho_A^*(T))=\frac{\sqrt{2} \rho_A^*(T)}{\sqrt{1 - \operatorname{Tr}(\rho_A^*(T)^2)}}.
\end{equation}

Together, the system dynamics, the adjoint equation, the optimal control law, and the transversality condition form a complete set of necessary conditions that must be satisfied by any optimal control trajectory. These conditions provide the foundation for computing optimal solutions in quantum entanglement generation via control strategies.

\section{Numerical Example: Optimal Entanglement Generation}
In this section, we numerically demonstrate the effectiveness of the proposed optimal control strategy for generating entanglement in a two-qubit system. The goal is to steer the quantum state toward a maximally entangled target by applying time-dependent control fields, as developed in the previous sections.

To this end, we consider a bipartite quantum system composed of two coupled qubits. The system evolves under unitary dynamics governed by the Liouville–von Neumann equation, with a total Hamiltonian consisting of both drift and control components. The Pauli matrices, which serve as the fundamental building blocks of the system Hamiltonians, are defined as:
$$\sigma_x = \left(
\begin{smallmatrix} 0 & 1 \\ 1 & 0 \end{smallmatrix}
\right), \quad 
\sigma_y = \left(
\begin{smallmatrix} 0 & -i \\ i & 0 \end{smallmatrix}
\right), \quad 
\sigma_z = \left(
\begin{smallmatrix} 1 & 0 \\ 0 & -1 \end{smallmatrix}
\right).$$
We consider a drift Hamiltonian \( H_d \) that represents cross-qubit coupling via asymmetric interaction terms:
\[
H_d = \sigma_x \otimes \sigma_z + \sigma_z \otimes \sigma_x,
\]
This drift Hamiltonian captures a nontrivial exchange of information between qubits, combining both local and interaction effects in a non-symmetric form.
The control Hamiltonian \( H_c(t) \) is defined as a linear combination of time-dependent control fields \( u_k(t) \) acting through three fixed control operators:
\[
\begin{aligned}
H_1 &= \sigma_x \otimes \sigma_y - \sigma_y \otimes \sigma_x, \\
H_2 &= \sigma_y \otimes \sigma_z - \sigma_z \otimes \sigma_y, \\
H_3 &= \sigma_z \otimes \sigma_x - \sigma_x \otimes \sigma_z.
\end{aligned}
\]
These control terms provide access to antisymmetric, cross-qubit interactions that are essential for generating complex entangled dynamics across the two-qubit system.

The optimal control fields \( u_k(t) \) are computed using the necessary conditions derived from Pontryagin’s Minimum Principle, with the objective of maximizing concurrence within a fixed time interval
to reach the final entangled state. In the following, we present simulation results for different initial states and demonstrate how the control strategy successfully drives the system toward maximally entangled Bell states. We consider a general initial state of the two-qubit system as a superposition:
\[
|\psi(t)\rangle = \alpha_{00} |00\rangle + \alpha_{01} |01\rangle + \alpha_{10} |10\rangle + \alpha_{11} |11\rangle,
\]
where \( \alpha_{ij} \in \mathbb{C} \) are complex coefficients. The corresponding density matrix is given by $\rho = |\psi\rangle\langle\psi|$.
The target states in our simulation are the Bell states \eqref{bellphi+}, \eqref{bells}, which serve as benchmarks for evaluating the performance of the control protocol.

To avoid initialization near unentangled configurations (which can cause flat gradients in optimization), we initialize the system close to a separable state with a small entangled perturbation:
$\rho_i = (1 - \varepsilon)\rho_{\text{sep}} + \varepsilon\, \delta\rho$,
where \( \rho_{\text{sep}} \) is a separable state and \( \delta\rho \) introduces a slight entanglement bias through the Bell states. This ensures that the initial state lies in a neighborhood well-suited to smooth convergence. 
In the following, we simulate the evolution of this initial state under the control law derived from PMP and evaluate the concurrence over time to verify the achievement of a maximally entangled target for several initial conditions. We consider a fixed time horizon from the initial time \( t_i = 0 \) to the final time \( T = 1 \). The control bounds are set to \( u_{\max} = 1 \), such that each control component \( u_k(t) \in \{-1, +1\} \). Additionally, we set the perturbation parameter to \( \varepsilon = 0.001 \).

To evaluate the performance of the proposed switching-based optimal control strategy, we consider two distinct initial configurations. These cases are chosen to assess the controller’s ability to transform weakly entangled or nearly separable states into highly entangled ones within a finite control horizon.

\paragraph*{Case 1 (Initialization Near \(|10\rangle\))}
In the first numerical experiment, the system is initialized with the following density matrix:
\[
\rho_0 = (1 - \varepsilon)\, |10\rangle\langle 10| + \varepsilon\, |\psi^+\rangle\langle \psi^+| =
\left(
\begin{smallmatrix}
0 & 0 & 0 & 0 \\
0 & \tfrac{\varepsilon}{2} & \tfrac{\varepsilon}{2} & 0 \\
0 & \tfrac{\varepsilon}{2} & 1 - \tfrac{\varepsilon}{2} & 0 \\
0 & 0 & 0 & 0
\end{smallmatrix}
\right),
\]
where \( |\psi^+\rangle = \tfrac{1}{\sqrt{2}}(|01\rangle + |10\rangle) \) is the Bell state. This configuration represents a slightly perturbed separable state, with coherence introduced between \(|01\rangle\) and \(|10\rangle\) through the small off-diagonal terms. As a result, the state becomes weakly entangled, making it a suitable candidate to test the convergence behavior of the control strategy.

\begin{figure}[t]
    \centering
    \includegraphics[width=0.95\linewidth]{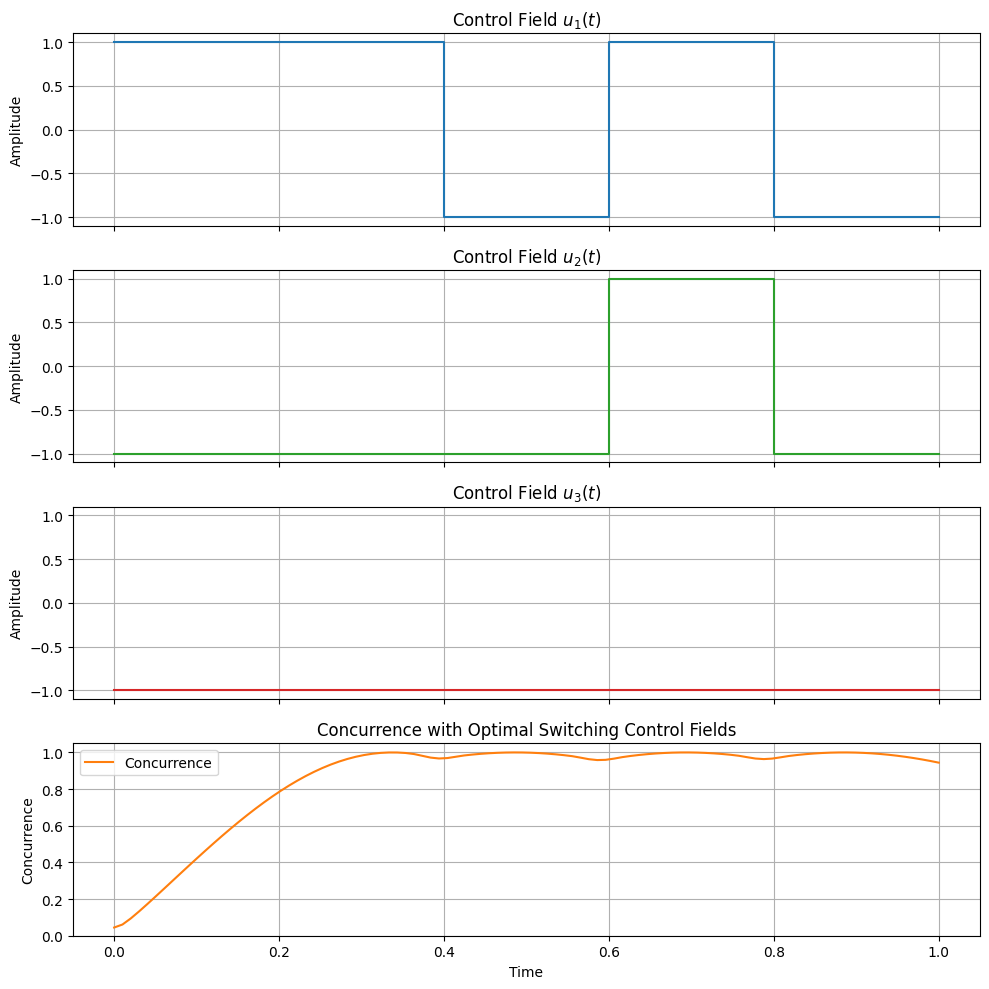}
    \caption{Time evolution of bang-bang control fields \( u_1(t), u_2(t), u_3(t) \) and the resulting concurrence for the initial state \( \rho_0 = (1 - \varepsilon) |10\rangle\langle 10| + \varepsilon |\psi^+\rangle\langle \psi^+| \). The control switching effectively steers the system to a highly entangled state.}
    \label{fig:optimal_concurrence_case1}
\end{figure}

\begin{figure}[t]
    \centering
    \includegraphics[width=0.95\linewidth]{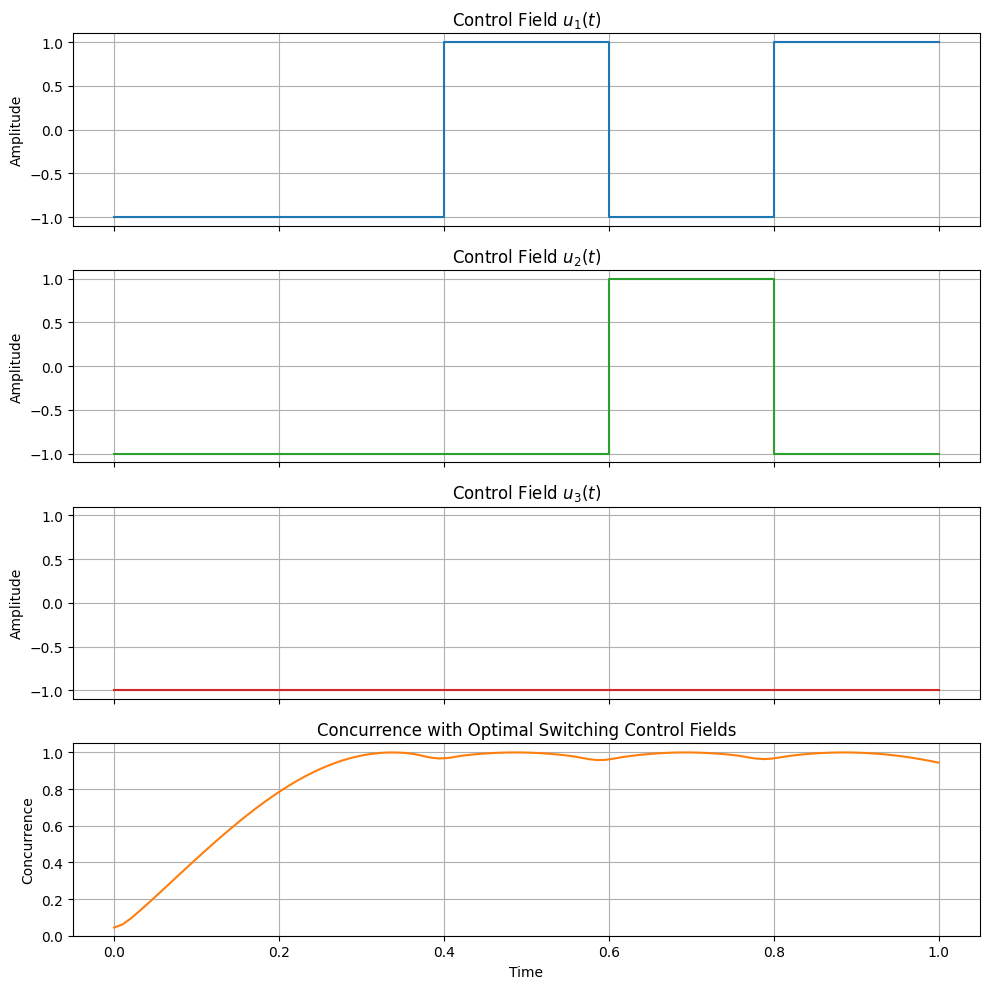}
    \caption{Optimally controlled entanglement dynamics for the initial state \( \rho_0 = (1 - \varepsilon) |01\rangle\langle 01| + \varepsilon |\psi^+\rangle\langle \psi^+| \). The top three plots show the bang-bang control fields \( u_1(t), u_2(t), u_3(t) \), and the bottom plot shows the resulting concurrence evolution.}
    \label{fig:01psiplus}
\end{figure}

Figure~\ref{fig:optimal_concurrence_case1} shows the time evolution of the optimized bang-bang control fields \( u_1(t), u_2(t), u_3(t) \), as well as the concurrence trajectory. The control inputs switch sharply between \(\pm u_{\max}\), consistent with Pontryagin’s Minimum Principle, and efficiently drive the system toward the entangled target. The concurrence increases steadily and saturates near 1, indicating that the system reaches a nearly maximally entangled configuration within the control horizon.
\paragraph*{Case 2 (Initialization Near \(|01\rangle\))}
In the second experiment, we initialize the system with:
\[
\rho_0 = (1 - \varepsilon)\, |01\rangle\langle 01| + \varepsilon\, |\psi^+\rangle\langle \psi^+| =
\left(
\begin{smallmatrix}
0 & 0 & 0 & 0 \\
0 & 1 - \tfrac{\varepsilon}{2} & \tfrac{\varepsilon}{2} & 0 \\
0 & \tfrac{\varepsilon}{2} & \tfrac{\varepsilon}{2} & 0 \\
0 & 0 & 0 & 0
\end{smallmatrix}
\right).
\]
This state follows the same structure as before, but with a different separable basis component. While the entangled perturbation is again \( |\psi^+\rangle \), the dominant weight is now on the \(|01\rangle\) level. This provides an additional perspective on the controller’s ability to handle different initial entanglement distributions. As shown in Fig.~\ref{fig:01psiplus}, the control fields retain their bang-bang structure, and the concurrence increases rapidly, eventually saturating near unity. This behavior confirms the robustness of the control strategy across different initializations and verifies its effectiveness in generating high-fidelity entanglement under bounded control inputs.


\paragraph*{Discussion} 
The numerical results confirm that the proposed control strategy reliably drives weakly entangled states toward maximal entanglement, with concurrence consistently reaching unity. The bang-bang structure of the control fields, as predicted by Pontryagin’s principle, ensures efficient and implementable dynamics. In both cases, entanglement builds up rapidly within the control horizon, demonstrating the method’s effectiveness.
\section{Conclusions}
This paper presented a quantum optimal control framework for robust entanglement generation in bipartite quantum systems. By formulating the problem using Pontryagin’s Minimum Principle, we derived a set of necessary optimality conditions to design time-dependent control fields that maximize entanglement, measured via concurrence, under bounded input constraints. Numerical simulations confirmed the validity and effectiveness of the approach. The results demonstrated fast and stable convergence of the concurrence to near-unity, highlighting both the robustness and practical viability of the method. The bang-bang nature of the optimal control inputs aligns with theoretical predictions and offers implementation advantages in realistic quantum hardware with amplitude limitations. This study contributes to the broader goal of controlled entanglement synthesis in quantum technologies and paves the way for future work on constrained quantum control in open-system settings, as well as extensions to multipartite or high-dimensional systems.

\bibliographystyle{IEEEtran}

\bibliography{ref}

\end{document}